\documentclass[jgrga,draft]{agutex2015}

\usepackage{float} 
\usepackage{auto-pst-pdf}
\usepackage{graphicx}
\usepackage{amssymb}

\authorrunninghead{ROBERTS ET AL.}

\titlerunninghead{ALFV\'EN VORTEX IN THE SLOW SOLAR WIND}

\authoraddr{Corresponding author: 
O. W. Roberts,
ESTEC, 
European Space Agency, 
Keplerlaan 1, 2201 AZ, Noordwijk, the Netherlands
Owen.Roberts@esa.int
}

\begin{document}

\title{Observation of an MHD Alfv\'en vortex in the slow solar wind}



\authors{
O. W. Roberts, \altaffilmark{1,2}
X. Li, \altaffilmark{2}
O. Alexandrova, \altaffilmark{3}
B. Li\altaffilmark{4}}

\altaffiltext{1}{ESTEC, European Space Agency, Keplerlaan 1, 2201 AZ, Noordwijk, the Netherlands}
\altaffiltext{2}{Department of Physics, Aberystwyth University, Aberystwyth, Ceredigion, SY23 3BZ, UK}
\altaffiltext{3}{LESIA, Observatoire de Paris, 5 place Jules Janssen, 92190 Meudon, France}
\altaffiltext{4}{Shandong Provincial Key Laboratory of Optical Astronomy and Solar-Terrestrial Environment,
    Institute of Space Sciences, Shandong University, Weihai, 264209, China}

%
%


\keypoints{
\item Fluctuations have low speed in the plasma frame indicative of an advected structure or a slowly propagating wave with $k_{\perp}\gg k_{\parallel}$
\item Comparison of data from the four Cluster spacecraft with an Alfv\'en vortex model shows excellent agreement 
\item Polarization of the fluctuations is consistent with a vortex structure
}


%
%


\begin{abstract}
In the solar wind, magnetic field power spectra usually show several power-laws. 
In this paper, magnetic field data from the Cluster mission during an undisturbed
    interval of slow solar wind is
    analyzed at 0.28~Hz, 
    near the spectral break point between the ion
    inertial and dissipation/dispersion ranges. 
Assuming Taylor's frozen-in condition, it corresponds to a proton kinetic scale of $kv_A/\Omega_p
\sim 0.38$, where $v_A$ and $\Omega_p$ are the Alfv\'en speed
and proton angular gyrofrequency, respectively. Data show that the
Cluster spacecraft passed through a series of wavepackets. A strong
isolated wavepacket is found to be in accordance with the four
Cluster satellites crossing an Alfv\'en vortex, a nonlinear solution
to the incompressible MHD equations. A strong agreement is seen between the data from four satellites and a model vortex with a radius of the order of $40$ times the local proton gyro-radii. The polarization at different spacecraft is compared and is found to agree with the vortex model, whereas it cannot be explained solely by the linear plane wave approach.
\end{abstract}

%
%

%

\begin{article}

\section{Introduction}

In neutral fluids, turbulence yields eddies from large to ever smaller scales until
   the turbulent energy is eventually dissipated by viscosity. 
In plasmas, the magnetic field brings complications that not only eddies but waves and current sheets
   are also commonplace, and  all these contribute to the dissipation of the turbulence power. 
Kinetic effects make studying the turbulence more challenging at ion and electron kinetic scales.

The solar wind is one of the best natural laboratories to
    study the plasma turbulence \citep{1995SSRv...73....1T, 2013LRSP...10....2B}. 
The existence of a magnetic field makes the solar wind turbulence highly
    anisotropic with $k_{\perp}\gg k_{\parallel}$ \citep{1983JPlPh..29..525S, 1996JGR...101.2511B, 1995ApJ...438..763G},
     where $k_\perp$ and $k_\parallel$ represent wavenumbers along directions
    perpendicular and parallel to the mean magnetic field, respectively.
 This anisotropy tends to be true at both MHD and ion kinetic
    scales \citep{Horbury2008,Podesta2009,Wicks2010,Chen2010,2010PhRvL.105m1101S,2011GeoRL..38.5101N,2013ApJ...769...58R, Roberts2014Thesis, 2015ApJ...802....2R}. 
 A typical magnetic field turbulent power spectrum involves an energy injection scale with
    a scaling of $k^{-1}$  for low wavenumbers, where Alfv\'enic turbulence dominates and energy is deposited into the system. 
 At intermediate wavenumbers, an ion inertial range with a $k^{-5/3}$ Kolmogorov scaling  is present
    until reaching a spectral break at ion scales ($k\rho_{i} \sim 1$ or $kd_{i}\sim 1$, 
    where $\rho_{i}$ and $d_{i}$ are the proton Larmor and inertial lengths, respectively).
The spectrum steepens beyond  this spectral break \citep{1998JGR...103.4775L, 2001JGR...106.8253S, 2006ApJ...645L..85S, 2008JGRA..113.1106H}. At scales smaller than  ion scales and up to electron scales, the spectrum follows a scaling of around -2.8
   \citep{2009PhRvL.103p5003A, 2010PhRvL.105m1101S, 2012ApJ...760..121A}.
At MHD scales,  turbulence is dominated by Alfv\'enic fluctuations \citep{1971ApJ...168..509B}.

 The nature of solar wind turbulence is still an open question: can waves still be used to describe turbulence (as a first approximation) or is it necessary to adopt the strong turbulence paradigm? To understand turbulent heating in space plasmas it is essential to understand the different contributions of these different phenomena to the overall energy budget. Dissipation in relation to waves may come from Landau damping or cyclotron resonance, while for coherent structures the possible mechanisms are reconnection or currents. Simulations by \cite{Karimabadi2013} and observations by \cite{2013ApJ...769...58R} suggest that coherent structures and waves may coexist in the solar wind. Therefore understanding which paradigm best describes the observed fluctuations has relevance for not only dissipative heating but also the turbulent cascade itself. Some properties of turbulence fluctuations such as megnetic helicity and, dispersion plots have often been interpreted in the wave paradigm as being due to kinetic Alf\'ven waves (KAWs) or a mixture of KAWs and ion cyclotron waves \citep{He2011,2015ApJ...802....1R}.
 Strong turbulence may be dominated by nonlinear coherent structures
     such as current sheets \citep{1968JGR....73...61S, 2007JGRA..11211102V}, 
     magnetic vortices like the Orszag-Tang vortex \citep{1979JFM....90..129O},
     or Alfv\'en vortices of the MHD type \citep{1985JETPL..42...54P, 1992swpa.book.....P}, drift type \cite{Shukla1985} or kinetic type \cite{Shukla1985a}.
 In a broader context, some detailed observations of coherent vortices are available in the Earth's and Saturn magnetic environments. 
Observational evidence of drift vortices in the Earth's ionosphere can be found in \citep{1988PhyS...38..841C, 1996JGR...10113335V}. 
Large-scale Kelvin-Helmholtz vortices have been observed on the Earth's magnetopause \citep{2004Natur.430..755H}. 
Kinetic Alfv\'en vortices were identified with multi-point Cluster measurements in the magnetospheric cusp region \citep{2005Natur.436..825S,Sundkvist2008}. 
While the first observational evidence of MHD Alfv\'en vortices in space plasmas was presented in \citet{2006JGRA..11112208A}, 
     where a multipoint analysis with Cluster clearly shows the topology of these magnetic structures
     and their propagation in the plasma frame. 
While these observations were made in the Earth's magnetosheath, 
     \citet{2008GeoRL..3515102A} showed the existence of such structures in the magnetosheath of Saturn.
 Regarding the solar wind, the only published signatures of vortex structures were presented by \citet{2003NPGeo..10..335V}
     using single satellite measurements, where a particular kind of polarization and discontinuities in the solar wind were explained with an Alfv\'en vortex model. 
More recently, a study by Lion et al. (2016) shows the presence of Alfv\'en-vortex-like structures 
     in the fast solar wind as measured with the Wind spacecraft.
These stuctures occur close to ion characteristic scales,     
     similar to what happens to the vortices observed
     in the Earth's magnetosheath \citep{2006JGRA..11112208A}.

 The studies of \citet{2008GeoRL..3515102A} and Lion et al.(2016) both employed single point measurements.
As such, they cannot definitively demonstrate the spatial localization of Alfv\'en vortices. 
A multi-satellite analysis is needed. 
 A recent statistical study of coherent structures around ion scales by Perrone et al. (2016, submitted to APJ) shows the presence of Alfv\'en vortex-like structures in a compressible slow wind stream. These structures have $k$ $\perp$ to $\mathbf{B_0}$ and slow propagation in the plasma rest frame, that was possible to estimate with four Cluster spacecraft. The space localization is verified, but the fluctuations have not been compared to the vortex model on four satellites to confirm the interpretation by the Alfv\'en vortex.
In two recent papers by \citet{2013ApJ...769...58R, 2015ApJ...802....2R}, 
     a k-filtering analysis based on 4 satellites measurements has shown that
     turbulent fluctuations around ion scales have $k_{\perp}\gg k_\parallel$
     and $\omega \simeq 0$ in the plasma frame. 
This was interpreted as a mixture of kinetic Alfv\'en waves (KAWs) and coherent structures such as vortices. 
\citet{2013ApJ...769...58R} also performed an analysis of the polarization of magnetic field fluctuations in the plane perpendicular
     to the global background magnetic field $\mathbf{B_{0}}$. 
In this plane, several coherent rotations of the magnetic field fluctuations were observed,
     indicating the presence of coherent structures. 
Here we re-analyze one of the time intervals examined in \citet{2015ApJ...802....2R}
     to show that it is possible to identify an Alfv\'en-vortex structure using simultaneous measurements from all four Cluster spacecraft. 
The end result is that we give clear
     evidence of the existence of an Alfv\'en vortex in the solar wind.

\section{Data and Methodology}

We use the magnetic field data obtained from the Fluxgate Magnetometer instrument (FGM) \citep{2001AnGeo..19.1207B} on the Cluster mission \citep{1997SSRv...79...11E}. 
    A ten minute interval which occurs on the $16^{th}$ of February 2005 between 22:30 and 22:40UT is studied, when the craft were in the slow solar wind.
The angle between the magnetic field and the bulk velocity is quite
    large ($\theta_{\mathbf{vB}}>60^{\circ}$), indicating that there is
    no magnetic connection to the bow
    shock.
 The E-field spectrogram from the WHISPER \citep{2001AnGeo..19.1241D} instrument is
    quiet (not shown), with no signatures of high frequency waves characteristic of the foreshock \citep{Lacombe1985,Alexandrova2013}. 
Some typical plasma parameters obtained from Cluster C1
    FGM and the Cluster Ion
    Spectrometer (CIS) \citep{2001AnGeo..19.1303R} are given in Table~\ref{pars1}.
 For this chosen event, the magnetic
    field is relatively stable and free from obvious discontinuities.
The latter is required because discontinuities
    would give large changes in $\mathbf{B_{0}}$ and $\delta B_{\parallel}$,
    thereby violating the incompressibility assumption of the vortex model. In addition to the low compressibility this interval was also selected since the Cluster spacecraft configuration were close to a regular tetrahedron, and the corresponding spatial scale of the wavepacket is larger than the interspacecraft distances ensuring that all spacecraft see the same wavepacket.
This interval was previously analyzed by \citet{2015ApJ...802....2R}  who concluded that the
    dispersion plot at scales slightly larger than those studied here
     ($kv_{A}/\Omega_{p} \sim 0.3$) were characteristic of either kinetic
    Alfv\'en waves or static structures. 
    
\begin{figure*}
\includegraphics[width=0.45\textwidth]{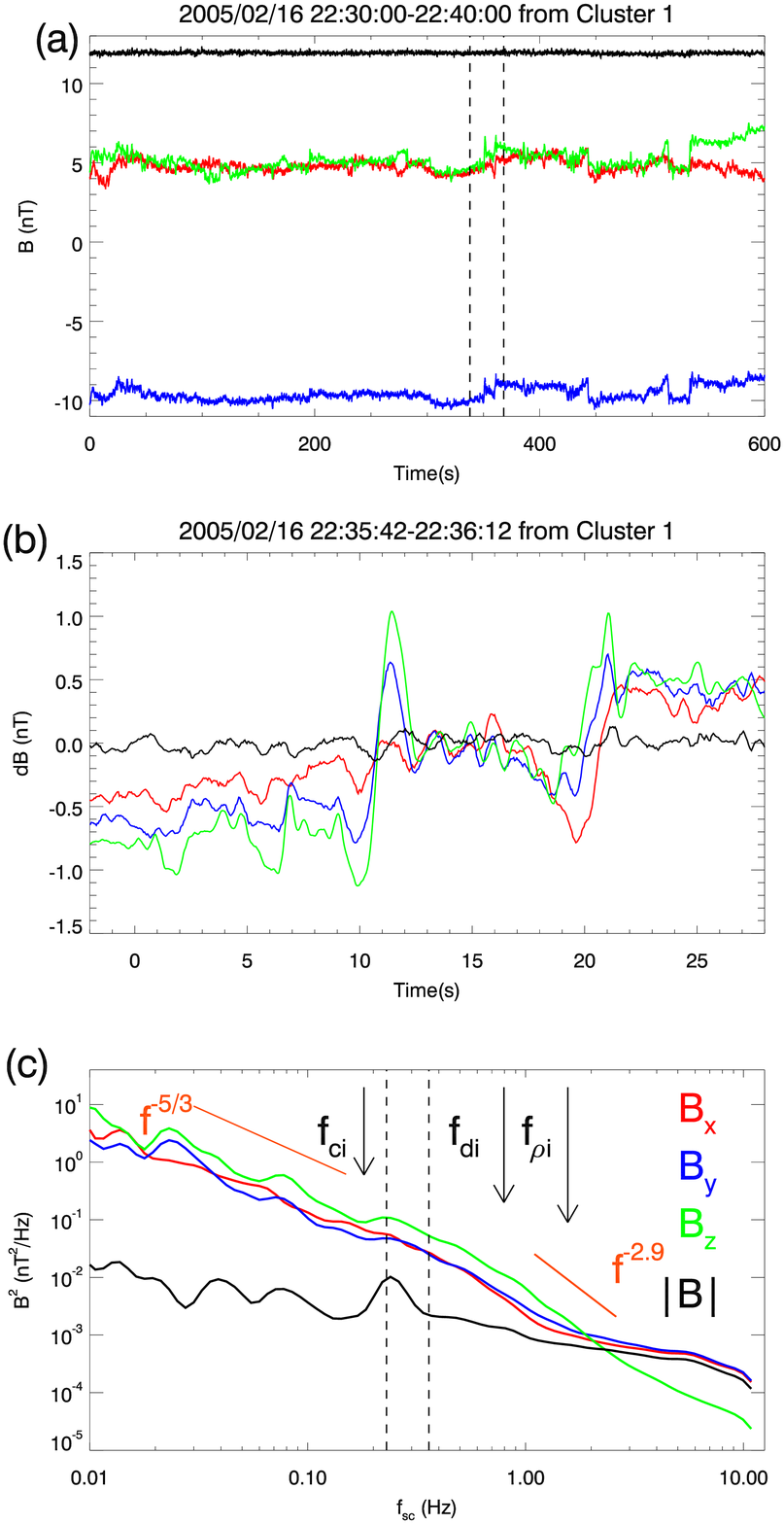}
\caption{
(a) Raw magnetic field time series in the GSE coordinates. 
Red, blue, and green denote the $x$-, $y$-, and $z$-components 
    of the magnetic field and black denotes the magnitude. 
(b) The fluctuations between the two vertical dashed lines magnified where the local mean (time average of 30 seconds) has been subtracted. The time series have been smoothed with a boxcar average with a width of 8 data points ($0.32$~s) in order to avoid the noise of the FGM instrument at $f\ge 3$~Hz., $t=0$s corresponds to 22:35:42UT.
(c) Wavelet power spectra of the magnetic field data in (a). The vertical lines in (c) denote the range of frequencies where
    bandpass filtering is performed, while the arrows denote the
    location of the cyclotron frequency and the 'Doppler-shifted'
    gyro-radius and inertial length observed at frequencies $f_{\rho i}=V_{sw}/2\pi \rho_{i}$ and  $f_{di}=V_{sw}/2\pi d_{i}$ respectively. Orange lines show power law scalings as a guide.} 
\label{tsps}
\end{figure*}

 Figure \ref{tsps}a  shows the raw magnetic field data in the Geocentric Solar Ecliptic (GSE) coordinate system from the FGM instrument. One can see that $|\mathbf{B}|$ does not vary much in the interval.
Figure \ref{tsps}b  shows the magnetic fluctuations defined here as $dB_{i}=B_{i}-\langle B_{i} \rangle_{30s}$ where the time average is done over the 30 seconds between 2 vertical lines of panel (a). Here one can see coherent, localized in time event in the middle of the time interval, between 5 and 15 seconds, visible mostly in $dB_{y}$ (blue) and $dB_{z}$ (green) components. At around 10.5 seconds all three fluctuation components are zero, suggesting that the spacecraft pass through a localized current sheet or a current filament at this point. At the end of the interval, there is another event with 3 components changing in phase. In our study we will focus on the central structure at $t=10$s.
Fig.1(c) shows the power spectral density (PSD) of the three magnetic field components and of $|\mathbf{B}|$ for the 10 minutes time interval shown in panel (a). One can see that between 0.2 and 0.3~Hz, there is an local maximum on the PSD($|\mathbf{B}|$), that can be a satellite spin effect (we will discuss this point in more details below).  In the PSD of the components, at the same frequencies one observes a spectral knee. Then, between 0.2 and 1~Hz, the $PSD(B_z)$ (green line) follows a clear power-law, which breaks to a steeper scaling of around $-2.9$ at $f>1$~Hz .Other magnetic field components arrive to the noise floor at $f>1$~Hz, so we can not conclude about the shape of the PSD of $B_x$ and $B_y$ at high frequencies It appears that the noise floor (where the spectra flatten due to instrumental noise near $f \gtrsim 1.5Hz$)  appears lower in the $B_{z}$ component compared to $B_x$ and $B_y$, the components in the satellite spin plane. If it was related only for the spin problem, it would be expected to be similar for all spacecraft, which is not observed for spacecraft C2 and C4 where the noise floor is similar for all three components (not shown).
. 


\begin{table*}
\caption{
Spacecraft and mean plasma parameters. 
Here $v_{sw}$ and $v_{A}$ denote the bulk speed and Alfv\'en speed, respectively.
In addition, $B_{0}$ is the magnitude of the magnetic field, 
    $n$ is the number density of protons, 
    $f_{ci}$ is the proton gyration frequency, 
    $\theta_{vB} (^{\circ})$ is the angle between the magnetic field and the bulk velocity, 
    $T_{i\perp}/T_{i\parallel}$ is the ratio of perpendicular to parallel temperatures of protons,
    and $d_{i}$ and $\rho_{i}$ denote the ion inertial length and Larmor radius, respectively. 
The plasma beta is denoted by $\beta$,
    and $d_{min}$ represents the minimum distance between a Cluster spacecraft pair.}
\begin{tabular}{ccccccccccc} \hline
\hline
$v_{sw}$		& $B_{0}$		&$v_{A}$ 				& $n$
			& $f_{ci}$ 		&$\theta_{\mathbf{vB}}$ 			& $T_{i\perp}/T_{i\parallel}$
			& $\rho_{i}$		&$d_{i}$				& $\beta$
			&$d_{min}$ 	\\
(km s$^{-1}$)		& (nT)			& (kms$^{-1}$)				& (cm$^{-3}$)
			& (Hz)			& $(^{\circ})$				&
			& (km)			& (km)					&
			&(km) \\
\hline			
377			&11.9			& 85.8 					&9.2
			& 0.181			& $112^{\circ}$				&0.5
			&38.4			&75.4					&0.57
			& 896.8\\
\hline
\label{pars1}
\end{tabular}
\end{table*}

Figure 2 shows magnetic scalogrammes calculated with Morlet wavelet transform for 3 components of magnetic field but in a primed coordinates, where $\mathbf{e}_{z'}$ is the unit vector along the background magnetic field $\mathbf{B_{0}}$, the other two unit vectors are defined as follows:
\begin{equation}
\centering \mathbf{e}_{x'}=[-B_{y}A/B_{x},A,0], \label{Prime2}
\end{equation}
\begin{equation}
\centering \mathbf{e}_{y'}=\mathbf{e}_{z'}\times \mathbf{e}_{x'},
\label{Prime3}
\end{equation}
where $A=B_{x}/\sqrt{B_{x}^{2}+B_{y}^{2}}$.
This system was chosen such that the velocity vector (predominantly in the $-x$ GSE direction) is mostly in the $+x'$ direction (since the largest component of $\mathbf{B_{0}}$ is in the negative y (GSE) direction see Fig. 1a).
The background magnetic field $\mathbf{B_{0}}$ used here is the global average for the ten minute interval.
It is important to note, that the local mean field around the structure shown in Fig.1b and the global mean (the mean for the full ten minute interval) are similar in this time interval.

In the plane perpendicular to $\mathbf{B_{0}}$ (see Fig.~2a and 2b), one observes localized energetic events covering a range of scales, from $\sim1$ to $\sim5$ seconds.Exactly, at the corresponding frequencies, $(0.2,1)$~Hz,  we observe a power-law spectrum in Fig.1c, as discussed above. 
These events have different energies and vary a bit in scales. In case of a technical issue, like spin, it would appear in a scalogram as a constant energy emission at a fixed scale, that is not the case of energetic events observed here.    Thus, there is no clear spin effect present in the scalograms 
 and spectral leakage of any fluctuations due to the spacecraft spin are not likely to affect the magnetic fluctuations of the localized energetic events.  

 The coherent magnetic fluctuations of Fig.1b corresponds to the energetic event between 2 vertical dotted lines in the scalogrammes: here the energetic peak appears around 3.6 seconds time scale. We will study magnetic fluctuations associated with this energetic event around its central scale, between frequencies $(0.23-0.36)$~Hz (between the dashed lines in Fig.~1c and Fig 2). For this purpose,  the data in the primed coordinate system are bandpass-filtered using a wavelet transform \citep{1998BAMS...79...61T} and  reconstructed as time series \citep{2013ApJ...769...58R} 
     such that only signals from this narrow range of frequencies are present. By assuming Taylor's frozen-in condition, the wavenumber sampled at
    $0.28$~Hz (the centre of the enhancement in Fig. 2b) has a component along the solar wind flow
    of $kv_A/\Omega_p \sim 0.38$.

\begin{figure}
\centering
\includegraphics[width=0.8\textwidth]{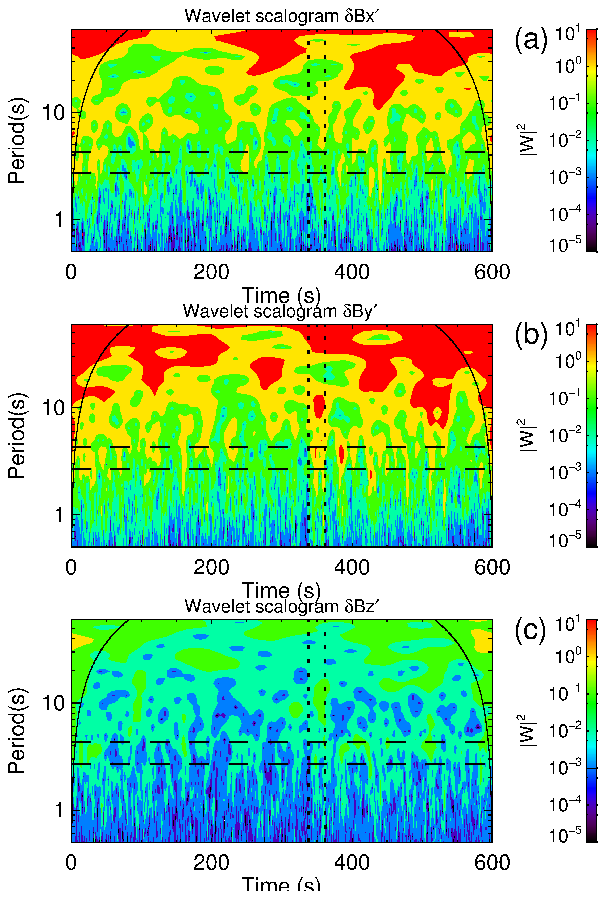}
\caption{Wavelet scaleograms showing the wavelet power for the fluctuations in the three primed coordinates. The solid line denoting the cone of influence region, the dotted lines denoting where we see the wavepacket, and the dashed lines denoting the region where we perform frequency filtering.} 
\label{scaleo}
\end{figure}

The reconstructed time series of the three components of magnetic field fluctuations
     are shown in Fig.3. 
These magnetic field fluctuations 
     are intermittent and consist of wavepackets.    
Here we are able to show that one such wavepacket is best described as an Alfv\'en vortex. 
The fluctuating magnetic field has very weak compressibility 
     since the parallel component (Fig. 3c) is
     substantially smaller than the two perpendicular components (Figs. 3a and 3b). 
We will focus on  an isolated wavepacket seen within the zoom boxes of Fig. 3 
    between $t=5$ and $15$ seconds. 
We find that the signals from C1 and C2 in the subinterval are stronger than from the other two satellites.

 The spatial scale of the wavepacket can be obtained
    without Taylor's hypothesis by using
    the phase differencing method \citep{1995GeoRL..22.2653D, 2004AnGeo..22.3021W}.  
It is important to note that this technique can only recover the wavenumber of the dominant
    fluctuation at a fixed spacecraft frequency for a wavepacket, and assumes that it can be described as a plane wave.
The difference in phase of the wavepacket as observed by two
    separate spacecraft can be estimated by using a cross correlation to
    measure the phase shift $\Delta \psi_{i,j}$ between the two signals
    at spacecraft pairs $i$ and $j$. 
This is related to the wavevector $\mathbf{k}$ by 
\begin{equation}
\Delta \psi_{i,j}=|\mathbf{k}||\mathbf{r_{i,j}}| \cos \theta_{\mathbf{kr}}
\label{phase}
\end{equation}
    where $\mathbf{r}$ is the
    separation vector between two spacecraft, $\theta_{\mathbf{kr}}$ is the
    angle between the wavevector and the spacecraft separation vectors.
Essentially $|\mathbf{k}|\cos \theta_{\mathbf{kr}}$ is the
    projection of the true wavevector $\mathbf{k}$ onto the spacecraft
    separation vector $\mathbf{r_{ij}}$. 
Cluster's four spacecraft give us the ability to compare
    the projected wavevector along three
    separate baselines thus allowing the determination of the true
    wavevector. 
The wavevector projections are related to the true wavevector via
\begin{equation}
\mathbf{k}\cdot\mathbf{A}=\mathbf{k'}~,
\label{proj1}
\end{equation}
     where $\mathbf{A}$ is a $3 \times 3$ matrix whose elements are given
     by three components of the unit vectors of the
     spacecraft separation vectors corresponding to the three projected
     wavevectors \citep{2003GeoRL..30.1508B}. 
These equations can be solved by inverting $\mathbf{A}$. 
For this wavepacket we investigate at a single scale corresponding central frequency of 0.28Hz where the enhancement is most intense in Fig. 2. Other scales of the energetic event yield similar results. The wavevector obtained for this wavepacket (from the $\delta B_{y'}$ component) makes a perpendicular angle
     with the global mean magnetic field $\theta_{\mathbf{k}\mathbf{B_{0}}}=90.2^{\circ}$. The wavenumbers in the direction parallel and perpendicular to the magnetic field are $k_{\parallel} \sim 1.8 \times 10^{-5}$~km$^{-1}(k_{\parallel}v_A/\Omega_p \sim 0.0008$) and  $k_{\perp} \sim 5\times 10^{-3}$~km$^{-1}$($k_{\perp}v_A/\Omega_p \sim 0.377$) respectively. The corresponding parallel and perpendicular scales are $\lambda_{\parallel}\sim350,000$km $(9000\rho_{i})$ and $\lambda_{\perp}\sim 1200$ km$ (31.3\rho_{i})$. The wavepacket can be seen in Fig~3 to have approximately 2.5 complete cycles,
     and therefore the diameter can be estimated to be 2.5 $\times 1200=3000$km ($78.1\rho_{i}$). 
 By Doppler shifting the frequency to the plasma frame ($\omega_{pla}=\omega_{sc}-\mathbf{k}\cdot\mathbf{v}$),
    a low frequency is obtained of $(0.06 \pm0.04) \Omega_{p}$,
    consistent with previous applications of the k-filtering method at these scales
    \citep{2010PhRvL.105m1101S, 2013ApJ...769...58R, 2015ApJ...802....2R}.
The error is calculated by assuming a $2.5\%$ error on the velocity for the duration of the wavepacket of the solar wind.
A low plasma frame frequency is indicative of either a slowly moving structure 
    (or one that is advected by the bulk flow) 
    or a linear Kinetic Alfv\'en wave. The corresponding phase speed estimated from this analysis $V_{ph \perp}=(1.8 \pm 1.4)$~km/s  or $(0.021 \pm 0.016) v_{A}$.
Note that using this method it is difficult to differentiate between these two scenarios
    based solely on phase differencing, given the error on the plasma frame frequency.
We will now show that the wavepacket is best  interpreted as an Alfv\'en vortex.

\begin{figure}
\centering
\includegraphics[width=0.8\textwidth]{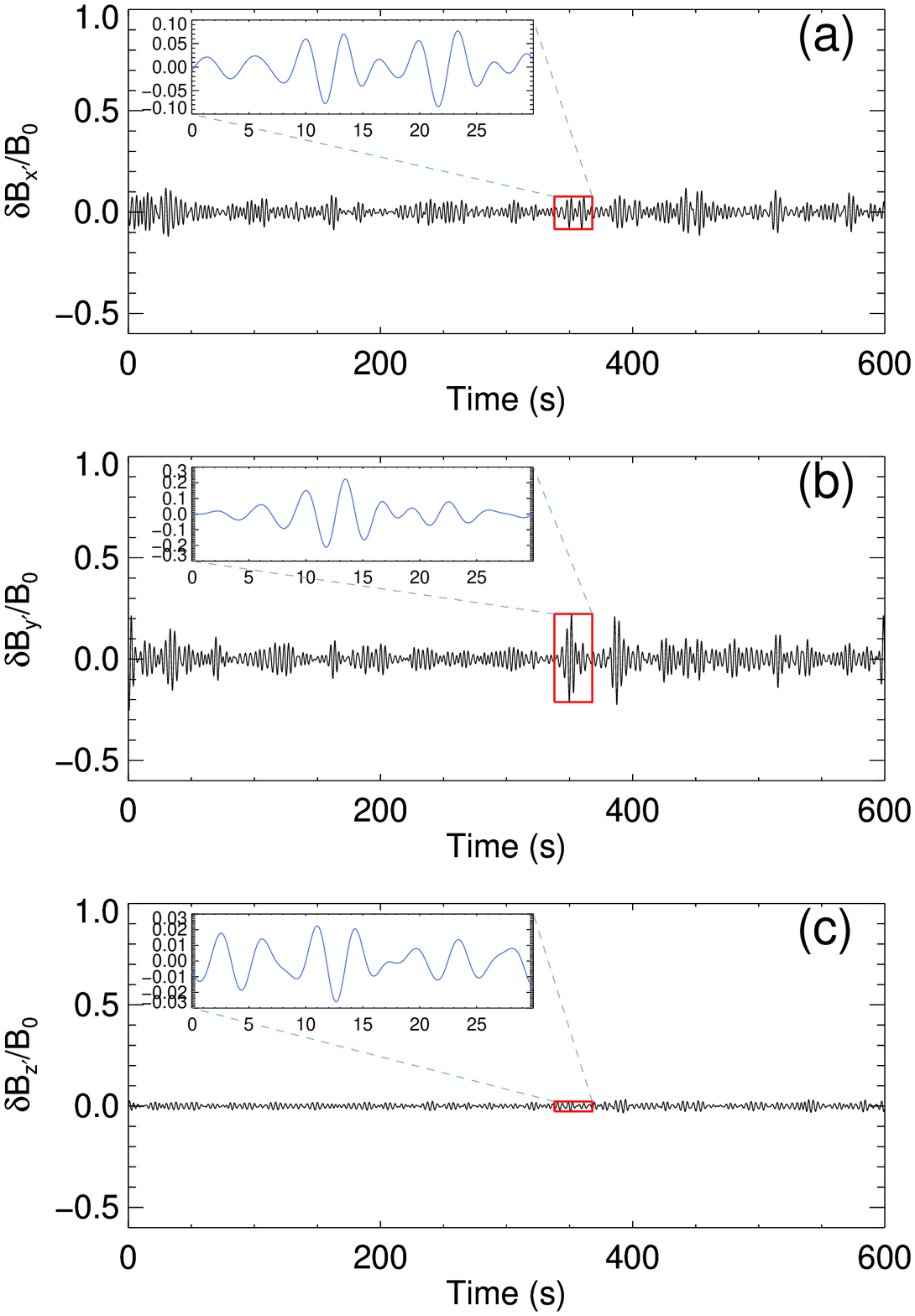}
\caption{Reconstructed time series of $\delta B_{x'}$ (a), $\delta
B_{y'}$ (b), and $\delta B_{z'}$ (c) within (0.23-0.36)~Hz frequency range. The wavepacket within the two
vertical dashed lines, which is one of the strongest in the time
interval, will be analyzed thoroughly next. The data are from
spacecraft C1.} 
\label{packet}
\end{figure}

\section{Alfv\'en Vortex Model}
Two types of MHD Alfv\'en vortex exist:  
    the monopolar one is perfectly aligned with $\mathbf{B_{0}}$
    ($\theta_{\mathrm {vortex}}$, the angle between the vortex axis and $\textbf{B}_0$, is $0^{\circ}$),
    whereas the dipolar one makes a small angle with $\mathbf{B_{0}}$
    ($\theta_{\mathrm{vortex}}>0^{\circ}$).

These vortices are tubular structures quasi-aligned with
$\mathbf{B_0}$ and are nonlinear solutions to the incompressible MHD
equations. They can be regarded as an MHD counterpart to
neutral fluid vortices and have been discussed theoretically by \cite{1985JETPL..42...54P, 1992swpa.book.....P}.

A dipolar vortex propagates with a small velocity $u$ in the direction of $y'$ relative to
     the plasma frame. 
It is convenient to describe the vortex solution using a variable
\begin{equation}
\eta = y'+\alpha z'-ut, ~~~ \alpha=\tan(\theta_{\mathrm{vortex}}),
\label{eqeta}
\end{equation}
     where $u= \alpha v_A$ is the speed of the vortex. 
The full derivation of the vortex can be found in \citep{1992swpa.book.....P} and \citep{2008NPGeo..15...95A},
     and we simply quote the results expressed with the $z'$-component of the vector potential
    ($\textbf{B}_\perp =\nabla A \times \textbf{e} _{z'}$).
The analytical solution  reads \citep{2008NPGeo..15...95A}
\begin{equation}
\left\{
\begin{array}{l}
A=A_{0}\left(J_{0}(kr)-J_{0}(ka)\right) 
    -\displaystyle\frac{2\alpha x'}{kr}\displaystyle\frac{J_{1}(kr)}{J_{0}(ka)}+\alpha x',~~~~r<a\\
A=A_{0}a^2\displaystyle\frac{\alpha x}{r^2},~~~~r\geq a ,
\end{array} \right.
\label{potA}
\end{equation}
 where $r=\sqrt{(x')^2+\eta^2}$, $A_0$ is a constant amplitude, 
     and $J_n (n=0, 1)$ is Bessel function of the $n$-th order.
 Furthermore, $a$ is the vortex radius. For continuity of the solutions, $ka$ must be one of zeros of Bessel function $J_1$. Here we will use the third zero of $J_1$, $ka=10.17$ to best model the three crests present in the principle fluctuation.  
 The resulting fluctuations within the vortex $r<a$ are then given by 
\begin{equation}
\left\{
\begin{array}{l}
 \delta B_{x'} =kA_0 J^{*}_0(kr) \displaystyle\frac{\eta}{r} \\ 
    + \displaystyle\frac{u}{k\xi}\displaystyle\frac{2}{J_0(ka)}
      \left[\displaystyle\frac{J_1(kr)}{r} -kJ^{*}_1(kr) \right]
      \displaystyle\frac{x'\eta}{r^2}, \\ \\
\delta B_{y'} =-kA_0 J^{*}_0(ka) \displaystyle\frac{x'}{r} \\
    + \displaystyle\frac{u }{k\xi}\displaystyle\frac{2}{J_0(ka)}
    \left[\displaystyle\frac{J_1(kr)}{r}\eta^2 +kJ^{*}_1(kr)(x')^2 \right]
          \displaystyle\frac{1}{r^2} -\displaystyle\frac{u}{\xi}.
\end{array} \right.
\label{dbxdby}
\end{equation}
 Here the starred $J_0$ and $J_1$ denote the derivatives respect to their arguments, 
    and $\xi=\frac{u}{\alpha}$ is a constant of order unity. 
For $r \geq  a$, we have
\begin{equation} 
\left\{ \begin{array}{l}
\delta B_{x'} = -\displaystyle\frac{2a^2 u}{\xi}\displaystyle\frac{x' \eta}{r^4} \\
\delta B_{y'} =  \displaystyle\frac{2a^2 u}{\xi} \left(\displaystyle\frac{(x')^2}{r^4}
     -\displaystyle\frac{1}{2r^2} \right).
\end{array}\right.
\label{dbxdby2}
\end{equation}

The magnetic field fluctuations seen by the spacecraft due to the vortex depend on several parameters, some of which are intrinsic to
     the vortex and some depend on the path of the spacecraft through the vortex. In the following section we will now describe these parameters of the specific vortex model that we observe.   
\section{Model Comparison with Solar wind data}

Figure~4d-g (solid lines) show magnetic fluctuations on the four Cluster satellites for the same wavepacket observed by C1 and shown in Figure 2. One can see that the four satellites observe equivalent signals but not at the same time. These fluctuations will now be compared to the fluctuations from the Alfv\'en vortex model, described above. 

In section 2 we have estimated using four satellites the spatial scale of the fluctuations which we can use here as vortex radius  $a=39.06\rho_{i} \sim 1500km$. We have also shown, that these fluctuations are convected (in the limits of the error) by the solar wind. Thus the only free parameters to fit will be: a single impact parameter, i.e.  the distance of a satellite path to the vortex center at $\eta=0$, in terms of its radius $a$;  an amplitude $A_{0}$, and the inclination of the vortex axis $\theta_{\mathrm{vortex}}$. It is important to note that an impact parameter may only be fitted for a single spacecraft, the impact parameters of other spacecraft are constrained by the relative distances to the first spacecraft.

\begin{figure*}[H]
\centering
\includegraphics[width=0.55\textwidth]{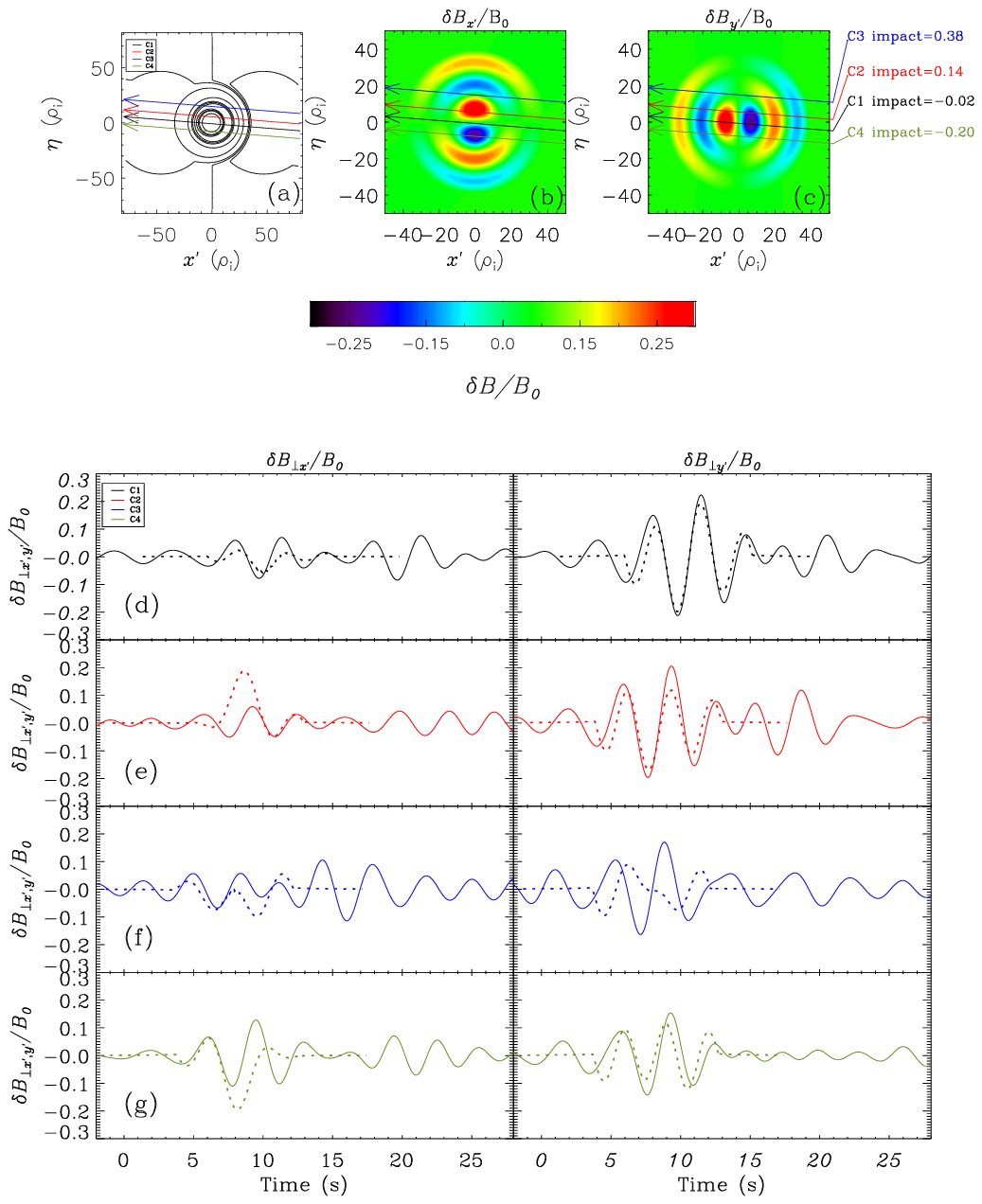}
\caption{Properties of a quasi monopolar Alfv\'en vortex, the axis of the vortex (which is defined as the normal to the $x'-\eta$ plane) makes an angle
of $0.35^{\circ}$ with $\mathbf{B_{0}}$. The
axes denote distance in units of $\rho_{i}$, this vortex propagates
at $0.006v_{A}$ relative to the bulk plasma. Figure (a) shows the
magnetic field lines. Figures (b) and (c) show the perpendicular
fluctuations due to the vortex. The arrows denote the paths of the
spacecraft through the vortex which give the modeled fluctuations in
Figs. (d-g). Lighter colours refer to positive $\delta B$, darker
colours refer to negative $\delta B$ as shown in the colour
bar. The impact parameters for the various spacecraft are given in
units of vortex radius and denote the distance from the vortex axis, passed by $(x',\eta)=0$. 
The spacecraft trajectories are denoted by arrows. Panels(d)-(g) show the observed fluctuations
(solid line) and the  modeled fluctuations (dashed lines) which
correspond to the trajectories presented in (a)-(c). The left panel
shows $\delta B_{x'}$ and the right panel shows the
$\delta B_{y'}$ with C1 at the top and C4 at the bottom.} 

\label{vmodel}
\end{figure*}

Figure~\ref{vmodel}a shows the vector potential from Eq.~(\ref{potA}), while Figures~4b and 4c show the resulting magnetic field fluctuations, $\delta B_{x'}$ and $\delta B_{y'}$, from Eqs. (\ref{dbxdby}) and (\ref{dbxdby2}), respectively. The arrows denote the trajectories taken by the spacecraft through the vortex.
In Figs.~\ref{vmodel}d-g we show the comparison of the data on four satellites with the Alfv\'en vortex model fluctuations (dashed lines) measured along the synthetic satellite trajectories shown in Figs. \ref{vmodel}a-c. 
Here we use  $A_{0}=-1.3 B_0 \rho^2_i \Omega_p /v_A$  and $\theta_{\mathrm{vortex}}=0.35^{\circ}$. Such small inclination corresponds to a very slow propagation speed of the vortex in the plane perpendicular to $\mathbf{B_0}$, namely  $0.006~v_{A}$.  The satellites paths are defined by the solar wind flow in the plane perpendicular to ${\bf B}_0$ and by the separations between the satellites, known a-priori. The minimal distance from the centre of the vortex to the path of C1 is determined to be $0.02a$, by varying the impact parameter and comparing the model fluctuations to the data until a good fit is found for the C1 craft. As well, we use  $ka=10.17$ (i.e. the third zero of Bessel function $J_1$), as mentioned above.

A strong agreement for all spacecraft between the measured signals and the modeled signals is seen in Figs. \ref{vmodel}d-g for the principal component, which is the $\delta B_{y'}$ component,  and to a lesser degree for $\delta B_{x'}$. An interesting result of the fitting is that the C1 craft pass close to the centre of the vortex where we would expect a current filament \cite{2008NPGeo..15...95A}, which corresponds exactly with the signature of a current seen in Fig. 1b. The model shows a stronger agreement for craft C1 and C2 than the other craft. This may be due to the fact that the C1 and C2 craft have smaller impact parameters making their trajectories closer to the centre of the vortex where the amplitudes of the fluctuations are larger. The fitting for $\delta B_{y}'$, for the C3 craft may be less accurate because of the larger impact parameter (see Fig.4c), and the spacecraft pass through a region where the amplitudes are smaller. Outside the vortex radius the analytical solution shows evanescant behaviour which is also seen near $t=2-7$s, however near $t=15-20$s data shows oscillatory behaviour with a smaller amplitude than within the vortex. This is due to the presence of another structure or wave in the vicinity (at $t=20$s Fig. 1b) of the studied vortex at $t=10$s in Fig 1b. Other discrepancies may also arise either from weak compressibility, or that kinetic effects are beginning to play a role at these scales.

\begin{figure}[H]
\centering
\includegraphics[width=0.4\textwidth]{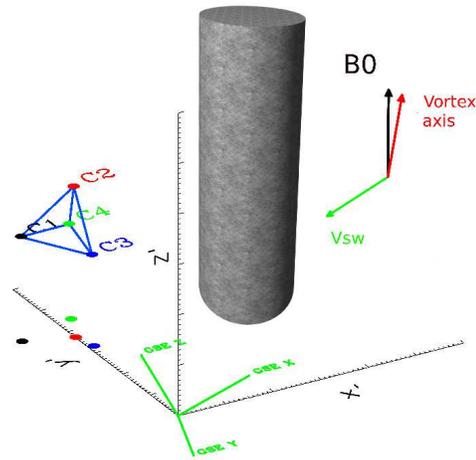}
\caption{
The Cluster spacecraft positions while crossing a tubular structure (Alfv\'en vortex). 
The four Cluster satellites are colour coded and are located at the vertexes of a tetrahedron. 
The cylindrical structure denotes the vortex inclined at a small angle
    to the mean magnetic field, which is directed along the $z'$-direction by construction.
The solid coloured dots are the projection of the satellites on the $x'Oy'$ plane.}
\label{vortBo}
\end{figure}

 Figure~\ref{vortBo} depicts the  scenario we observe here: an Alfv\'en vortex is crossed by the Cluster spacecraft.
 The projections of the spacecraft positions onto the $x'Oy'$ plane are shown by the coloured dots.

In our case, the path of the spacecraft makes an angle with the axis of the vortex of $112^{\circ}$, since the magnetic field and solar wind flow form such an angle.  
Since the displacement of the spacecraft in the $z'$ direction is not constant, this will increase the effective radius seen by the spacecraft. In our case in this coordinate system, the increase of the radius seen by the spacecraft is small ($\sim200$km) compared to the size of the vortex ($\sim3000$km), and the fluctuations do not vary along the $z'$ axis.

The vortex axis is indicated here by a red arrow. From the fitting, we have determined that the angle $\theta_{\mathrm{vortex}}=0.35^{\circ}$.
Note that this angle is exaggerated in the figure for presentation purposes. 
 Strictly speaking, a monopolar vortex is perfectly aligned with the magnetic field
     direction and is advected with the solar wind bulk flow. 
As the angle between the magnetic field direction increases the vortex becomes a dipolar one, 
     and a quadrupolar structure \citep[][Fig.~3]{2008NPGeo..15...95A} 
     is seen in the perpendicular magnetic field components. 
However, since the angle is so small a quadrupolar structure
     is not seen here. 
We will use the term `quasi monopolar' to describe this vortex.

\begin{figure*}[H]
\centering
\includegraphics[width=0.98\textwidth]{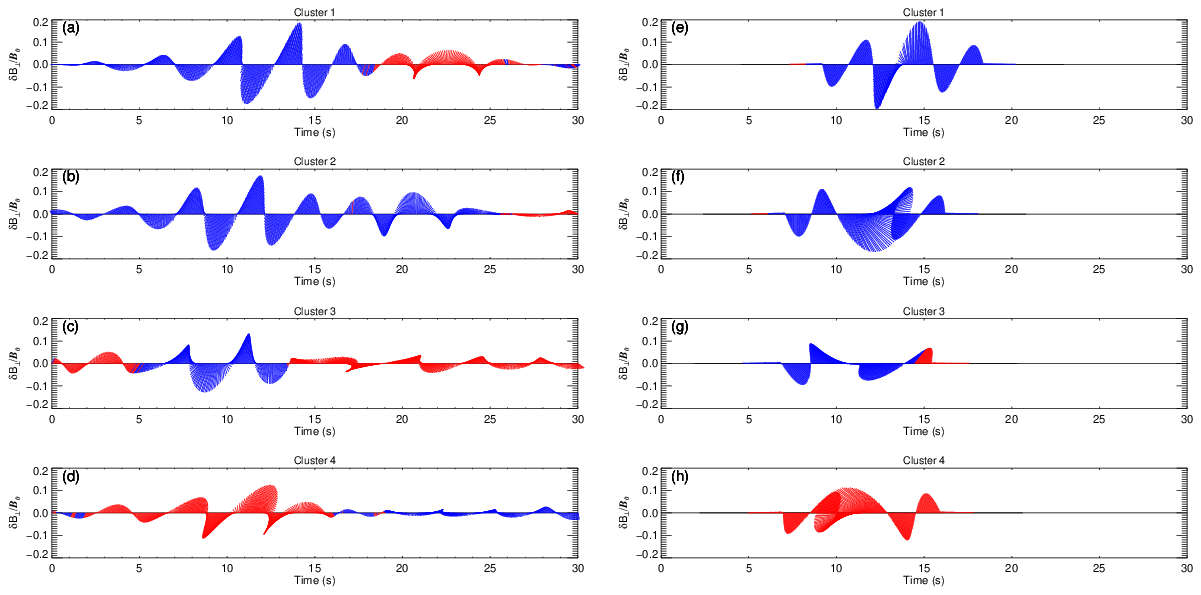}
\caption{Left panels (a-d): Comparison of the polarization observed from the spacecraft ; 
Right panels, (e-h): the polarization of the Alfv\'en vortex model fluctuations, shown by dashed lines in Fig.3(d-g).}
\label{polar}
\end{figure*}

Polarization analysis is another diagnostic technique that can be
    used to investigate the fluctuations. 
Here we consider the polarization in the plane perpendicular
    to the global mean magnetic field $\mathbf{B_{0}}$. 
In this plane coherent rotations may signify the presence of coherent structures \citep{1996JGR...10113335V}. 
Additionally, the sense of polarization would depend on the path of the spacecraft through the vortex, 
    which could be left/right handed or linear. 
 Figures \ref{polar}a-d show this sense of polarization with red (blue) lines denoting right (left) handed
    sense of rotation.   
 Note that for the wave interpretation we would expect the
     polarization to have the same sense for the same wavepacket regardless of the point of observation.
However, we see that two spacecraft show a mix of both senses of polarization (C1, C2),
    and the remaining two spacecraft show a strong sense of rotation in opposite directions (C3, C4).
These fluctuations are compared to the polarizations predicted from
    the vortex model in Figs. \ref{polar}e-h. 
Good agreement is found between the hodographs obtained from data and the model. 
A curious feature is that the hodographs vary between spacecraft, 
    and are close to being linearly polarized for C1 and C2 but polarization is
    opposite between C3 and C4, which is not expected for a plane wave.
The vortex interpretation, however, can explain this rather satisfactorily (see the right panels of Figure~5). 
An alternative explanation is that the spacecraft were observing two different wavepackets.
 However, this seems unlikely since cross correlation analysis gives
    high values for the similarity of the signals between spacecraft,
    with the largest similarity being $0.96$ and the smallest being $0.84$
    for the $B_{y'}$ components. 
 Thus the wavepacket presented here cannot be described by the wave paradigm, 
    while the Alfv\'en vortex model reproduces nicely the fluctuations and the polarization
    for all spacecraft.

\section{Conclusion}
To summarize, the Cluster spacecraft offer a unique opportunity
    to study plasma turbulence in three dimensions. 
We have discussed both linear wave and nonlinear structure paradigms in relation to turbulence. 
While they are two very different concepts, their measured signatures are very similar
    and differentiating between both concepts is difficult, emphasising the need for multi-point measurements. 
Comparisons between the data obtained from the spacecraft and the Alfv\'en vortex model 
    show excellent agreement with the real spacecraft distances consistent with their distances in the model. 
We have also presented a study of the polarization, which shows features that cannot be
    explained using only linear wave formalism.

In conclusion, we have presented clear evidence of a quasi-monopolar Alfv\'en vortex in the solar wind.
For the wavepacket concerned here, a coherent structure aligned with
    the magnetic field explains the data consistently while the linear Alfv\'en wave
    interpretation alone cannot fully describe the observations. 
Further research is needed to study whether and how such Alfv\'en vortices are
    involved in the turbulent cascading process in the solar wind at 1AU.
%
%
%
%
%
%
%

\begin{acknowledgments}
 All Cluster data are obtained from the ESA Cluster Science Archive.
We thank the FGM and CIS instrument teams and the CSA team . OR acknowledges helpful discussions with Khurom Kiyani and Chris Carr. This study is supported in part by the National Natural
Science Foundation of China (40904047 and 41174154).
\end{acknowledgments}

\end{article}
%
%
%
%
%
%
%
%


\end{document}